\documentclass[pra,twocolumn]{revtex4}
\usepackage{epsfig}
\usepackage{psfig}
\usepackage{latexsym,exscale}
\usepackage{bm}
\usepackage{amsmath,amssymb,amsfonts}
\def\be{\begin{eqnarray}} 
\def\ee{\end{eqnarray}}

\def\bp{\bm{p}}

\def\bP{\bm{P}}
\def\bx{\bm{x}}
\def\bxp{\bm{x'}}
\newcommand{\Fd}{F^{\dagger}}

\newcommand{\1}{{\uparrow}}
\newcommand{\2}{{\downarrow}}

\begin{document}
\title{Pairing in a two-component ultracold Fermi gas: phases with
  broken space symmetries}
\author{Armen Sedrakian$^1$, Jordi Mur-Petit$^2$, Artur Polls$^2$ and Herbert M\"uther$^1$}
\affiliation{$^1$Institute for Theoretical Physics, 
T\"ubingen University, D-72076 T\"ubingen, Germany\\
$^2$Departament d'Estructura i Constituents de la Mat\`eria,
Universitat de Barcelona, E-08028 Barcelona, Spain
}


\begin{abstract}
We explore the phase diagram of a two-component ultracold atomic Fermi gas
interacting with zero-range forces in the limit of weak coupling. We
focus on the dependence of the pairing gap and the free energy
on the variations in the number densities of the two species while
the total density of the system is held fixed.
As the density asymmetry is increased, the system  exhibits 
a transition from a homogenous Bardeen-Cooper-Schrieffer (BCS) 
phase to phases with spontaneously broken global
space symmetries. One such realization is the 
deformed Fermi surface superfluidity (DFS) which 
exploits the possibility of deforming the Fermi surfaces
of the species into ellipsoidal 
form at zero total momentum of Cooper pairs.  
The critical asymmetries at which the transition from DFS
to the unpaired state occurs are larger than those for the BCS phase. 
In this precritical region the DFS phase lowers 
the pairing energy of the asymmetric BCS state. We compare quantitatively 
the DFS phase to another realization of superconducting phases 
with broken translational symmetry: the single-plane-wave 
Larkin-Ovchinnikov-Fulde-Ferrell
phase, which is characterized  by a nonvanishing center-of-mass momentum 
of the Cooper pairs. The possibility of the detection of 
the DFS phase in the time-of-flight experiments is discussed and 
quantified for the case of $^6$Li atoms trapped 
in two different hyperfine states.
\end{abstract}
\maketitle

\section{Introduction}

The progress achieved in recent years in trapping and manipulating 
ultracold fermion gases has focused much theoretical attention on the 
pairing properties of dilute fermionic systems. Current experiments 
with $^6$Li and  $^{40}$K atoms are carried out at temperatures
which are a fraction ($\sim 0.1-0.3$) of the 
Fermi temperature~\cite{MBEC1,MBEC2,MBEC3,MBEC4,MBEC5,MBEC6,MBEC7,SALOMON1}.
These systems are thus characterized by a filled Fermi sea, and 
at sufficiently low temperatures attractive two-body 
interactions are expected to drive the Cooper pairing instability~\cite{STOOF1,STOOF2}.
The strength of the two-body interactions can be tuned using the Feshbach 
resonance mechanism by varying the external 
magnetic field~\cite{FESHBACH1,FESHBACH2,FESHBACH3};
thus, the entire range from weak to strong
couplings can be probed. In the crossover region the Feshbach
resonance may strongly enhance the pairing interaction and give rise
to high-temperature  superfluidity~\cite{RP1,RP2,RP3,RP4,RP5,RP6,RP7,RP8,RP9,RP10,RP11,RP12}.
Recent experiments have probed the condensation of fermionic pairs above the 
Feshbach resonance, where the system does not support a genuine two-body
bound state~\cite{JILA04,MIT04,NC04}. The 
nature of these pairs (tightly bound molecules versus extended
Cooper pairs) is not clear yet, and the signatures 
of the superfluid phase transition, which are manifest in the 
collective and hydrodynamic behavior of these 
systems  \cite{BARANOV,STRINGARI}, have been searched for. 
The measured collective modes of $^6$Li
were found to be consistent with superfluid
hydrodynamics and provide some evidence for superfluidity 
in a resonantly interacting Fermi gas \cite{NC04}. A gap in the 
quasiparticle spectrum of a two-component gas of $^6$Li was recently 
observed using radio-frequency spectroscopy~\cite{grimm04}.

The $s$-wave interaction dominates the pairing interaction in cases
 where (i) the pairing is among atoms of equal mass number but
belonging to different 
hyperfine states or (ii) the pairing is between different species 
(in mixtures of atoms- e. g,. $^6$Li and $^{40}$K),
in which case the symmetric BCS limit is not realized~\cite{COMMENT}.
Systems where two hyperfine levels are populated
have been created and studied experimentally with $^6$Li and $^{40}$K
atoms~\cite{MBEC1,MBEC2,MBEC3,MBEC4,MBEC5,MBEC6,MBEC7,SALOMON1,JILA04,MIT04,NC04,SALOMON2,SALOMON3}. 
These systems are characterized by a hierarchy of scales: the typical
range of the van der Waals forces is $R\le 10^{-6}$ cm while the 
de Broglie wavelength of particles at the top of the Fermi sea is 
$k_F \sim 10^3$-$10^4$ cm$^{-1}$. Since $k_FR\ll 1$ 
the interaction can be approximated by a zero-range force 
which is characterized by the $s$-wave scattering length $a_S$.
We specify our discussion below to the case where two hyperfine states 
of $^6$Li are populated, in which case  the scattering length
in units of the Bohr radius is  $a_S/a_B = -2160$.
For typical values of 
Fermi momenta quoted above $k_Fa_S\simeq 0.04$ and 
the system is in the weak-coupling regime, since $\nu(k_F) U_0
 = (2/\pi) k_F\vert a_S \vert \ll 1$, 
where $\nu(k_F) = mk_F/(2\pi^2\hbar^2)$ is 
the density of states at the Fermi surface and $U_0 = 4\pi\hbar^2 a_S/m$
is the strength of the contact interaction. For larger values of $k_Fa_S
\sim 1$ the weight of the negative-energy states to the 
single-particle spectral function is not negligible - the bound states 
need to be incorporated in the theory along with the pair correlations 
on the same footing~\cite{levin}.

The BCS theory predicts a suppression of the pairing correlations 
when the Fermi energies or, equivalently, the densities 
of the two hyperfine states $\vert 1\rangle$ and $\vert 2\rangle$,
denoted below as $\rho_{\1/\2}$, are
different. In the low-density and weak-coupling limit 
($k_F|a_S| \ll 1$) the value of the critical
density  asymmetry $\alpha = (\rho_{\1}-\rho_{\2})/(\rho_{\1}+\rho_{\2})$, 
for which the superfluidity vanishes, follows from the 
relation~\cite{MUR,LOMBARDO}
\be
  \frac{\Delta (\alpha)}{\Delta_0} 
   = \sqrt{1-\frac{4\alpha}{3}\frac{\mu}{\Delta_0}} ,
\ee
where 
$\Delta_0\simeq {8 e^{-2}}\mu\exp{\left[-{\pi / (2k_F|a_S|})\right]}$ 
is the gap in the symmetric system and $\mu$ is the chemical potential.
Therefore, the gap disappears for asymmetries 
$\alpha >\alpha_{\rm c}^{BCS}=3\Delta_0/(4\mu)$. For example, if 
the pairing is between $^6$Li atoms in the states 
$|1\rangle=|F=3/2,m_F=3/2\rangle$, and
$|2\rangle=|3/2,1/2\rangle$, in which case the pairing interaction 
is characterized by a scattering length $a_S/a_B=-2160$, the maximum 
asymmetry at which BCS pairing is possible is $\alpha_{\rm c}^{BCS}
\simeq 0.055$ 
for a density $\rho=\rho_{\1}+\rho_{\2}=3.8\times10^{12}$ 
cm$^{-3}$ (corresponding to $k_F|a_S|=0.55$).

One purpose of this work is to show that the pairing correlations
in ultracold atomic gases can persist for density asymmetries 
$\alpha > \alpha_{\rm c}^{BCS}$ and can be enhanced for
$\alpha < \alpha_{\rm  c}^{BCS}$ if  the Fermi spheres of two hyperfine
states  are deformed into ellipsoids in momentum space. In the 
strong-coupling regime ($\Delta_0/\mu \sim 0.3$) the superconducting state featuring 
deformed Fermi surfaces was found preferable to the spherically 
symmetric BCS state~\cite{DFS_PRL}.  Here we rather explore the 
weak-coupling regime. Alternatively, the homogeneous 
BCS phase can evolve into the Larkin-Ovchinnikov-Fulde-Ferrell 
(LOFF) phase~\cite{LARKIN,FULDE}, which again 
sustains $\alpha > \alpha_{\rm c}^{BCS}$
asymmetries by allowing the Cooper pairs to carry finite center-of-mass 
momentum (the Fermi surfaces in the LOFF phase are spherical). 

There has been much interest in the LOFF phase outside the 
condensed matter context in connection with  hadronic systems 
under extreme conditions where the interactions are mediated by the 
strong force (see Ref.~\cite{QCD_LOFF}), but no experimental signature(s) 
of its realization has been proposed so far.
Atomic systems offer a novel setting for studying the LOFF phase under
conditions that are more favorable than those 
in solids (absence of the lattice  effects, access to the 
momentum distribution in the system through time-of-flight
experiments)~\cite{COMBESCOT,COMBESCOT_MORA1,COMBESCOT_MORA2,MIZUSHIMA,URBAN}. 
Atomic systems also offer the possibility of novel realizations of 
the LOFF phase which, for example, invoke $P$-wave anisotropic 
interactions~\cite{COMBESCOT} or finite-size systems~\cite{MIZUSHIMA,URBAN}.

While the LOFF and DFS phases break global space symmetries, 
the way they do so is  fundamentally different: the LOFF phase breaks 
both the rotational and translational symmetries due to 
the finite momentum of the condensate and 
irrespective of the form of the lattice structure; 
the DFS phase breaks only the rotational symmetry from O(3) down to O(2).
We shall compare below the realizations of the 
DFS and LOFF phases in the $^6$Li gas assuming 
that the order parameter in the LOFF phase has a simple plane-wave form;
in addition, we shall describe an experimental signature of the DFS phase
that can be established in time-of-flight experiments and 
that would allow one to distinguish the DFS phase from the 
competing phases. A brief account of this argument was given earlier 
in Ref.~\cite{DFS_SHORT}.
Yet another possible alternative is the phase 
separation of the superconducting and normal phases in real space,
such that  the superconducting phase contains particles with the same 
chemical potentials - i.e. is symmetric - while the normal phase
remains asymmetric; see Refs. \cite{Bedaque:1999nu,BCR,CALDAS}. A comparison 
of the heterogeneous phase with the alternatives would require knowledge 
of the poorly known surface tension between superconducting and normal phases, 
and will not be attempted here.

This paper is organized as follows. 
Starting from Dyson's equations, we derive in Sect.~II the dispersion 
relations of elementary excitations in an asymmetrical superfluid. 
Sect.~III discusses the gap equation and its
regularization for contact interactions. Sect.~IV is devoted to
modifications of the single particle spectra needed 
to describe the LOFF and DFS phases. Numerical results for the case 
of  $^6$Li gas are presented in Sect.~V. Sect.~VI discusses 
the possibility of observing the DFS phase in time-of-flight experiments.

\section{Formalism}
\begin{widetext}
Consider a uniform gas of fermionic atoms with two hyperfine states, which 
we assign labels $\1$ and $\2$ (these states equivalently can be thought of as 
pseudospins.). The model Hamiltonian that describes our system is 
\begin{eqnarray}\label{HAMILTON}
\hat H &=& \frac{1}{2m}
\sum_{\alpha}\int\! d^3 x
{\nabla} \hat \psi^{\dagger}_{\alpha}(\bx)
{\nabla} \hat \psi_{\alpha}(\bx) 
-\sum_{\alpha\beta}\int\! d^3 x\int\! d^3 x' \hat \psi^{\dagger}_{\alpha}
(\bx)  \hat \psi^{\dagger}_{\beta}(\bxp) V(\bx,\bx')
\hat \psi_{\beta}(\bxp) \hat\psi_{\alpha}(\bx),
\end{eqnarray}
where $\hat \psi^{\dagger}_{\alpha}(\bx)$ and $\hat \psi_{\alpha}(\bx)$ are the
creation and annihilation operators of a state at the space point specified 
by the position vector $\bx$ and pseudospin $\alpha (= \uparrow,\downarrow)$,
$m$ is the atom's bare mass, and  $V(\bx,\bx')$ is the two-body potential 
(here and below the volume $\Omega =1$). The one-body propagator in the
superfluid state is a $2\times 2$ matrix in Gor'kov space,
\be 
\underline G (x,x') =
\left( \begin{array}{cc}
 G_{\alpha\beta}(x,x') & F_{\alpha\beta}(x,x')\\
-F^{\dagger}_{\alpha\beta}(x,x') & \overline G_{\alpha\beta}(x,x')\\
\end{array}
\right) = 
\left( \begin{array}{cc}
-i\langle T\psi_{\alpha}(x)\psi_{\beta}^{\dagger}(x')\rangle 
&\langle \psi_{\alpha}(x)\psi_{\beta}(x')\rangle \\
\langle \psi_{\alpha}^{\dagger}(x)\psi_{\beta}^{\dagger}(x')\rangle
&-i\langle \tilde T\psi_{\alpha}(x)\psi_{\beta}^{\dagger}(x')\rangle\\
\end{array}
\right),
\ee
where $x = (\bx, t)$, the Greek indices stand for the pseudospin, $T$ and
$\tilde T$ are the time-ordering and time-antiordering symbols, and
$G_{\alpha\beta}(x,x')$ and $F^{\dagger}_{\alpha\beta}(x,x')$ are
the normal and anomalous propagators~\cite{abri}. The propagators
above are assumed to be ordered on the Schwinger-Keldysh real-time 
contour~\cite{KELDYSH}, which  permits a finite-temperature treatment of
the problem. To keep our presentation concise, we shall not write down 
the explicit form of the correlation functions on the contour; 
our final expressions are written for retarded correlation functions.  
The $2\times 2$ matrix Green's function satisfies the 
familiar Dyson equation
\be\label{DYSON}
\underline{  G}_{\alpha\beta}(x,x') = 
\underline{  G}^0_{\alpha\beta}(x,x') 
+ \sum_{\gamma , \delta}\int\!\!d^4x'' d^4x'''\underline{  G}^0_{\alpha\gamma}(x,x''')
\underline{ \Sigma}_{\gamma\delta}(x''',x'') 
\underline{  G}_{\delta\beta} (x'',x'),
\ee
where the free propagators $\underline{  G}^0_{\alpha\beta}(x,x')$ 
are diagonal in Gor'kov space; the underline indicates that 
the propagators and self-energies are matrices in this space.
The components of the Fourier transform of Eq. (\ref{DYSON}) with respect to 
the difference of the space arguments of the two-point correlation 
functions are
\be\label{1}
  G_{\alpha\beta}(p) &=&   G_{0\alpha\beta}(p) +  
G_{0\alpha\gamma}(p) \left[  \Sigma_{\gamma\delta}(p)   G_{\delta\beta}(p)
+  \Delta_{\gamma\delta}(p) \Fd_{\delta\beta}(p) \right],\\
\label{2}
 \Fd_{\alpha\beta} (p) &=&   G_{0\alpha\gamma}(-p)\left[  
\Delta^{\dagger}_{\gamma\delta}(p)   G_{\delta\beta}(p) 
+  \Sigma_{\gamma\delta}(-p)  \Fd_{\delta\beta} (p) \right],
\ee\end{widetext}
where $p$ is the four-momentum, $G_{0\alpha\beta}(p)$ is the 
free normal propagator, and 
$  \Sigma_{\alpha\beta}(p)$ and $  \Delta_{\alpha\beta}(p)$  
are the normal and anomalous self-energies;  
summation over repeated indices is understood. 
The Dyson equations for the components 
$\overline G_{\alpha\beta}(p)$ and $F_{\alpha\beta}(p)$ follow
from Eqs. (\ref{1}) and (\ref{2}) through the time-reversal operation.

Below, we shall assume that the interactions conserve spin- i.e.,
$  G_{\alpha\beta}(p) = \delta_{\alpha\beta}  G(p)$ 
and $ \Sigma_{\alpha\beta} (p)= \delta_{\alpha\beta} \Sigma(p)$
 - and concentrate on the pairing in the state of zero total spin 
and orbital angular momentum
$S = L = 0$. Thus the anomalous propagators and self-energies 
must be antisymmetric with respect to the spin indices,
\be
   \Fd_{\alpha\beta} (p) = g_{\alpha\beta}    \Fd (p), \quad 
 \Delta^{\dagger}_{\alpha\beta} (p) = g_{\alpha\beta} \Delta^{\dagger} (p), 
\ee
where $g_{\alpha\beta} \equiv i\sigma_y$ is the spin matrix,
with  $\sigma_y$ being the second component of the vector of Pauli matrices. 
It is convenient to rewrite Eqs. (\ref{1}) and (\ref{2}) in terms of  
auxiliary Green's functions, which describe the unpaired state of 
the system:
\be \label{N1}
  G^N_{\alpha\beta}(p) =   G_{0\alpha\beta}(p)+ 
G^N_{\alpha\gamma}(p) \Sigma_{\gamma\delta}(p)  
G^N_{0\delta\beta}(p).
\ee
The formal solution of Eq.  (\ref{N1}) for the retarded propagators
in terms of the self-energy $\Sigma(p)$ is
\be\label{NORMAL}
G^{N}_{\alpha\beta}(\pm p)
&=&\delta_{\alpha\beta}[\pm\omega-\xi_{p\sigma} -\Sigma_{\sigma}(\pm p)]^{-1}\nonumber\\ 
&=&\delta_{\alpha\beta}[\pm(\omega +i\eta)-\varepsilon _{p\sigma}
]^{-1}+O({\rm Im}\Sigma_{\sigma}(\pm p)) , \nonumber\\
\ee
where $\xi_{p\sigma} = p^2/2m_{\sigma} - \mu_{\sigma}$ is the energy of 
a free spin-$\sigma$ particle relative to the chemical potential $\mu_{\sigma}$; 
the second line follows in the quasiparticle approximation, which keeps the 
leading-order term in the expansion of the propagator with respect to the 
small imaginary part of the self-energy.
The quasiparticle dispersion relation in the normal state is then 
given by 
$\varepsilon_{p\sigma} = \xi_{p\sigma}+{\rm Re}\Sigma_{\sigma}(\pm p)$.
Combining Eqs. (\ref{1}) and (\ref{2}) with Eq. (\ref{N1}) one finds 
\be\label{D1}
  G_{\alpha\beta}(p) &=&   G^N_{\alpha\gamma}(p)
\left[\delta_{\gamma\beta}+ \Delta_{\gamma\delta}(p)
 \Fd_{\delta\beta}(p)\right],\\
\label{D2}
 \Fd_{\alpha\beta}(p) &=&   G^N_{\alpha\gamma}(-p) 
 \Delta^{\dagger}_{\gamma\delta}(p)   G_{\delta\beta}(p);
\ee
these equations are easily solved to obtain the propagators
\be 
G_{\1\2}(p) &=& \frac{\omega +E_S\pm E_A}{(\omega- E_A)^2-E_S^2-\Delta^2},\\
\Fd(p) &=& -\frac{\Delta^{\dagger}}{(\omega-E_A)^2-E_S^2-\Delta^2},
\ee
where $E_S = (\varepsilon_{p\1}+\varepsilon_{p\2})/2$ and 
$E_A = (\varepsilon_{p\1}-\varepsilon_{p\2})/2$ are, respectively, 
the parts of the spectrum which are 
symmetric and antisymmetric under time-reversal operation.
The poles of these propagators (which must be identical) 
define the dispersion relation of the quasiparticles 
in the paired state:
\be\label{SPECTRUM} 
\omega_{1/2}  = E_A \pm \sqrt{E_S^2+\Delta^2}.
\ee
Note that the quasiparticle spectrum is twofold split due to the
asymmetry in the number of the spin-up and spin-down atoms. In the
symmetric limit $(E_A = 0)$ we recover the ordinary BCS dispersion 
relation. 

\section{gap equation}
\label{SEC:GAP}
To obtain a closed set of equations we need to specify the approximation 
to the anomalous self-energy. In the mean-field (BCS) approximation,
\be \label{GAP}
\Delta^{\dagger}(p) &=&  i\int V(\bp,\bp') \Fd (p') \frac{d^4p'}{(2\pi)^4}
\nonumber\\
&=&  i\int V(\bp,\bp') G^N_{\2}(-p')\Delta^{\dagger}(p') G_{\1}(p')
\frac{d^4p'}{(2\pi)^4},\nonumber\\
\ee
where $V(\bp,\bp')$ is the Fourier transform of 
the two-body interaction $V(\bx,\bx')$ which is responsible 
for the pairing.  The pairing interaction can 
be renormalized in such a way that integration in Eq. (\ref{GAP}) 
is restricted to the vicinity of the Fermi surface. This permits us
to approximate smooth functions of momentum by their value
at the Fermi momentum. We introduce a momentum renormalization 
scale $\Lambda$ such that $\Delta \ll \varepsilon_{\Lambda} 
\ll {\rm min} [\varepsilon_{F\1},~\varepsilon_{F\2}]$, where 
$\varepsilon_{F\1\2}$ are the Fermi energies of spin-up and down 
species. Then,
\begin{widetext}
\be \label{GAP1}
\Delta^{\dagger}(p) &=& i\int
~U (\bp,\bp')
G^N_{\2}(-p')\Delta^{\dagger}(p')G_{\1}(p')
\theta (\Lambda-\vert\bp'\vert)\frac{d^4p'}{(2\pi)^4},\\
\label{RENORM}
U(\bp,\bp') &=& V(\bp,\bp')+i\int V(\bp,\bp'')
G^N_{\2}(-p'')G^N_{\1}(p'') U(\bp'',\bp')\theta (\vert\bp''\vert-\Lambda )
\frac{d^4p''}{(2\pi)^4}.
\ee
In the second equation the full propagator $G_{\1}(p)$ was replaced 
by its counterpart in the unpaired state, $G^N_{\1}(p)$.
By construction, the on-shell integration is carried 
over momenta much larger than the Fermi momentum for which the
quasiparticle spectrum is unaffected by the pairing gap.
Thus Eq.~(\ref{RENORM}) decouples from Eq.~(\ref{GAP1}), 
while the integration in Eq.~(\ref{GAP1}) is now 
constrained to the vicinity of the Fermi surface. 
This permits us to simplify the problem by 
expanding the pairing interaction in spherical harmonics with respect
to the angle formed by the momenta lying on the Fermi surface(s) and,
thus, to reduce the pairing interaction to a function of a single angle.
While this is useful for finite-range interactions, in the case
of zero-range interactions the potential $V(\bp,\bp')$ needs to be eliminated 
from Eq. (\ref{RENORM}) in favor of the scattering $T$ matrix, which 
obeys the integral equation 
\be
\label{TMAT}
T(\bp,\bp') &=& V(\bp,\bp')+i\int_0^{\infty} V(\bp,\bp'')
G^N_{\2}(-p'')G^N_{\1}(p'') T(\bp'',\bp')
\frac{d^4p''}{(2\pi)^4}.
\ee
Combining Eqs. (\ref{RENORM}) and (\ref{TMAT}) one finds a regular
integral equation defining the pairing force
\be
U(\bp,\bp') &=& T(\bp,\bp')- i\int_0^{\Lambda} U(\bp,\bp'')
G^N_{\2}(-p'')G^N_{\1}(p'') T(\bp'',\bp')
\frac{d^4p''}{(2\pi)^4}.
\ee
\end{widetext}
In the dilute limit of interest,
partial waves higher than the $s$-wave can be neglected, 
and the interaction is solely determined by
the $s$ wave scattering length $a_S<0$, as  
$ T(\bp,\bp') = T_0= 4\pi\vert a_S\vert/m$. The solution 
of Eq. (\ref{RENORM}) with this interaction is straightforward, and one
obtains for $ U(\bp,\bp') = U_0$

\be  \label{U}
U_0 = {T_0}\left[1-\int G^N_{\2}(-p)G^N_{\1}(p)
\theta (\Lambda -\vert\bp\vert )\frac{d^4p}{(2\pi)^4}\right]^{-1}.
\ee
For the zero-range interaction above the gap equation takes the form 
\be\label{GAP2} 
\Delta^{\dagger}(\bp) &=& U_0 \int {\rm Im } \Fd(\bp',\omega')f(\omega')
\theta (\Lambda-\vert\bp'\vert)\frac{d^3p'd\omega'}{(2\pi)^4}\nonumber\\
 &=& \frac{U_0}{2} \int_0^{\Lambda} \frac{\Delta}{\sqrt{E_S^2+\Delta^2}}
\langle f(\omega_1)-f(\omega_2) \rangle
\frac{p'^2dp'}{(2\pi)^2},\nonumber\\
\ee
where  $\langle \dots\rangle $ stands for the angle average
and the second line follows in the quasiparticle approximation
[i.~e., by retaining only the pole part of the  propagator $\Fd(p)$].
The densities of the species  are given by  
\begin{align}\label{OCCUP} 
\rho_{\1/\2} &= \int {\rm Im } G_{\1\1/\2\2}(\bp',\omega)f(\omega)
\frac{d^3p'd\omega'}{(2\pi)^4}\nonumber\\
&= \int \langle u(\bp')^2\left[f(\omega_{\1/\2})-f(-\omega_{\2/\1})\right]
+ f(-\omega_{\2/\1})\rangle \frac{p'^2dp'}{(2\pi)^2},
\nonumber\\
\end{align}
where $u(\bp)^2 = 1/2+E_S /(2\sqrt{E_S^2+\Delta^2})$ 
is the familiar Bogolyubov amplitude and the second line of
Eq. (\ref{OCCUP}) is obtained in the quasiparticle approximation.
The normal self-energy in the $T$-matrix (ladder) approximation is
defined as 
\be \label{SIGMA}
\Sigma_{\sigma}(\bp) &=&  i\sum_{\sigma'}
\int\! T_{\sigma\sigma'}(\bp,\bp';\varepsilon_{\bp}+\omega') 
G_{\sigma'}(p') \frac{d^4p'}{(2\pi)^4},\nonumber\\
\ee
where the dependence of the self-energy on the center-of-mass momentum 
is suppressed. If the $T$ matrix is approximated as above by a real constant
$T_0$ which accounts for interactions between different species, 
the normal self-energy is momentum independent, purely real, and is given by 
\be \label{SIGMA2}
\Sigma_{\1/\2} =  T_0 \rho_{\2/\1}.
\ee
This constant shift in quasiparticle energy can be absorbed in the 
chemical potential by defining $\mu_{\1/\2}^* = \mu_{\1/\2} -
\Sigma_{\2/\1}$. It is straightforward to extended this result to the 
case where the $T$ matrix contains components which act among the same 
species or includes contributions from partial waves with  $L>0$.
Equations (\ref{U}), (\ref{GAP2}), (\ref{OCCUP}), and (\ref{SIGMA2}) 
are the fundamental equations of the mean-field theory 
for systems with unequal spin pairing that are
interacting via zero-range forces.
They include implicitly  the effects of a finite momentum 
of Cooper pairs and the topology of the Fermi surfaces to which we
shall turn in the following section. 

\section{Finite pair momentum and Deformed Fermi surfaces}

While the BCS ground state assumes that fermions bound in a Cooper pair
have equal and opposite momenta (and spins), 
for fermionic systems with and unequal number
of spin-up and -down particles this is not always true. Larkin and Ovchinnikov
\cite{LARKIN} and independently Fulde and Ferrell \cite{FULDE} observed that 
pairing is possible among pairs which have a finite total momentum with 
respect to some fixed reference frame.  The finite momentum $\bP$ 
changes  the quasiparticle spectrum of the paired state. To see this 
note that the auxiliary propagator (\ref{NORMAL}) written in the 
center-of-mass frame reads
\be\label{NORMAL_CM}
G^{N}_{\1/\2}(p,\bP)
=\left[\omega 
-\frac{1}{2m}\left(\frac{\bP}{2}\pm \bp\right)^2 -\mu^*_{\1/\2}+i\eta\right]^{-1} ,
\ee
and, therefore, the symmetric and antisymmetric (under time-reversal)
parts of the quasiparticle spectrum are
\be 
E_S &=& \frac{1}{2}
\left(\frac{P^2+4p^2}{4m} - \mu_{\1}^*-\mu_{\2}^*\right),\\
E_A &=& \frac{1}{2}
\left(\frac{\bP\cdot\bp}{m} - \mu_{\1}^*+\mu_{\2}^*\right).
\ee
The results of Sec. \ref{SEC:GAP} remain valid with the above 
redefinitions of $E_S$ and $E_A$. Note that 
the quantities of interest, in particular the gap, 
now depend parametrically on the total momentum. As a consequence,
the twofold splitting of the spectrum (\ref{SPECTRUM}) 
does not vanish in the limit of an equal number of spin-up and 
-down particles (i.e., when $\mu^*_{\1}=\mu^*_{\2}$). While such
a state lowers the energy of the system with respect to 
the unpaired state, it is still
unstable with respect to the ordinary BCS ground state.

We now turn to the deformations of the Fermi surfaces. 
The two Fermi surfaces for spin-up and -down 
particles are defined in momentum space by
the equations $\varepsilon_{\1/\2} = 
\epsilon_{\bP/2\pm\bp} -\mu^*_{\1/\2} = 0$. When the states are
filled isotropically within a sphere, the chemical potentials 
are related to the Fermi momentum $p_{F, \sigma}$ 
as $\mu^*_{\sigma} = p^2_{F, \sigma}/2m$
(for the sake of argument we assume here that the temperature is zero).
To describe the deformations of Fermi surfaces from their spherical 
shape we expand the quasiparticle spectrum in spherical harmonics
$\varepsilon_{\sigma} = \sum_l \varepsilon_{l \sigma} P_l(x)$, where $x$
is the cosine of the angle formed by the particle momentum 
and a randomly chosen symmetry-breaking axis and
$P_l(x)$ are the Legendre polynomials. 
The $l = 1$ terms break the translational symmetry
by shifting the Fermi surfaces without deforming them; these terms are 
ignored below. Truncating the expansion at second order ($l =
2$), we rewrite the spectrum in a form equivalent to the above one \cite{DFS_PRL}
\begin{equation}
\varepsilon_{\1/\2} = \epsilon_{\bP\pm\bp}-
\mu^*_{\1/\2}\left(1+\eta_{\1/\2} x^2\right) ,
\end{equation}
where the parameters $\eta_{\sigma}$ describe the quadrupole deformation of the
Fermi surfaces. It is convenient to work with the symmetrized
$\Xi = (\eta_{\1}+\eta_{\2})/2$ and
anti-symmetrized $\delta\epsilon  = (\eta_{\1}-\eta_{\2})/2$ 
combinations of $\eta_{\1,\2}$. Below we shall assume $\Xi = 0$
and consider two limiting  cases  $\delta\epsilon\neq 0$ and  $\bP= 0$
(the phase referred as the deformed Fermi surface superfluidity
(DFS) phase) and $\delta\epsilon=0$  and $\bP\neq  0$ (the plane-wave LOFF phase).  

\section{RESULTS}

Consider a trap loaded with $^6$Li atoms and assume that the net number 
of atoms in the trap is fixed while the system is maintained at constant 
temperature. Assume further that the number of atoms corresponds to a 
Fermi temperature $T_F=\epsilon_F/k_B=942$ nK, which in the uniform, 
symmetric case at $T=0$ would translate into a Fermi momentum of the system
$k_F\approx4.83\times10^4$ cm$^{-1}$ and a density 
$\rho=3.8\times10^{12}$ cm$^{-3}$.
We shall work below at constant  temperature $T=10$ nK ($\ll T_F$);
i.e., the system is in a highly degenerate regime. 
(In the  experiments of Ref.~\cite{hadzi}, such a Fermi temperature 
corresponds to about 4$\times10^5$ atoms in a single hyperfine 
component.) Experiments control  
the partial densities of atoms in two different hyperfine states
 $|\1\rangle=|F=3/2, ~m_F=3/2\rangle$ and $|\2\rangle=|3/2, ~1/2\rangle$,
e.~g. by transferring atoms from one state to the
other using $\sim$76 MHz rf pulses~\cite{pulses}. 
Since the free-space triplet scattering length for $^6$Li atoms 
in these hyperfine states is $a_S =-2160a_B$, the
system is in the weakly coupled regime  $k_F\vert a_S\vert \ll \pi/2$.

\begin{figure}
\epsfig{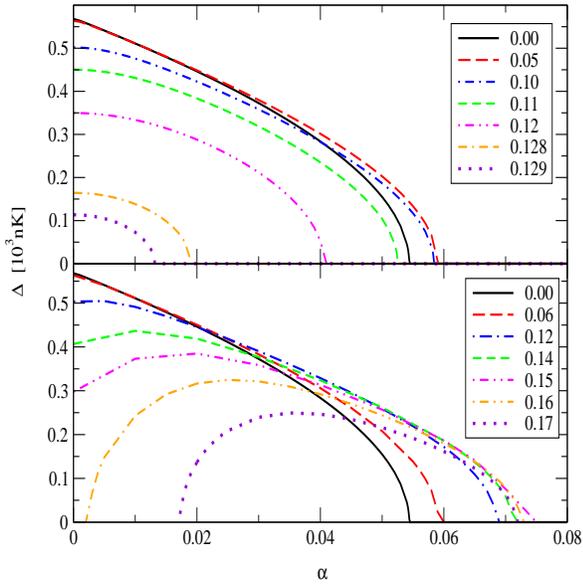}
\caption{  (Color online)
  The dependence of the pairing gaps in the LOFF phase (upper panel) 
  and the DFS phase (lower panel) on the asymmetry 
  parameter for several values of the the total momentum $P/k_F$
  and deformation parameter 
  $\delta\epsilon$   which are indicated 
  in the panels.  }
 \label{MSfig:fig1}
\end{figure}

The pairing gaps of the LOFF and DFS phases
computed from the coupled equations (\ref{GAP2}) and (\ref{OCCUP})
are shown as a function of asymmetry
parameter $\alpha = (\rho_{\1}-\rho_{\2})
/(\rho_{\1}+\rho_{\2})$  for different values of 
the total momentum $P$ and deformation $\delta\epsilon$
in Fig.~\ref{MSfig:fig1}. Without loss of generality 
the density asymmetry is constrained to positive values; i.e., we assume 
$\rho_{\1} > \rho_{\2}$. The positive values of $\delta\epsilon$ correspond 
to a prolate (cigar like) deformation of the majority and oblate
(pancake like) deformation of the minority population's Fermi spheres; 
we shall confine ourselves to the  case where $\delta\epsilon \ge 0$
since we have checked that it is the one corresponding 
to the largest value of the pairing energy.
\begin{figure}[tbh]
\epsfig{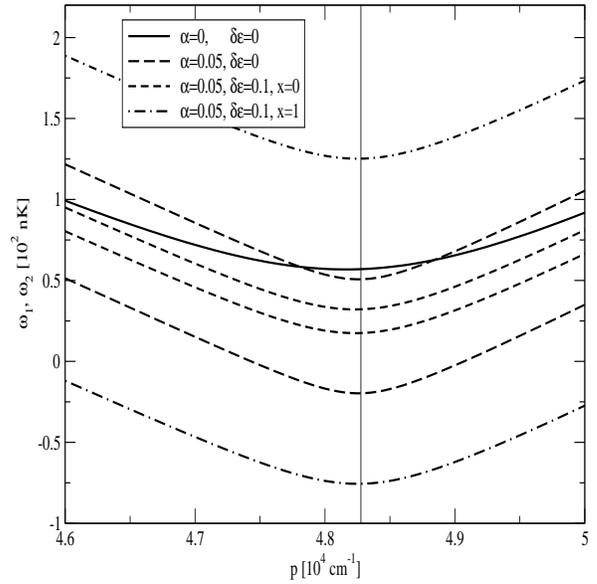}
\caption{
The dependence of the quasiparticle  spectra of two hyperfine 
states $\omega_1$ and $\omega_2$ on the momentum for $\alpha = 0=
\delta\epsilon$ (solid line),  
$\alpha = 0.05$ and $\delta\epsilon = 0$ (long-dashed lines),  
$\alpha = 0.05$, and $\delta\epsilon = 0.1$, $x = 0$ (short-dashed lines), 
and $x = 1$ (dash-dotted lines).
The Fermi momentum $k_F = 4.83\times 10^4$ cm$^{-1}$ is indicated 
by the vertical line.
}
\label{MSfig:fig2}
\end{figure}
\begin{figure}[tbh]
\epsfig{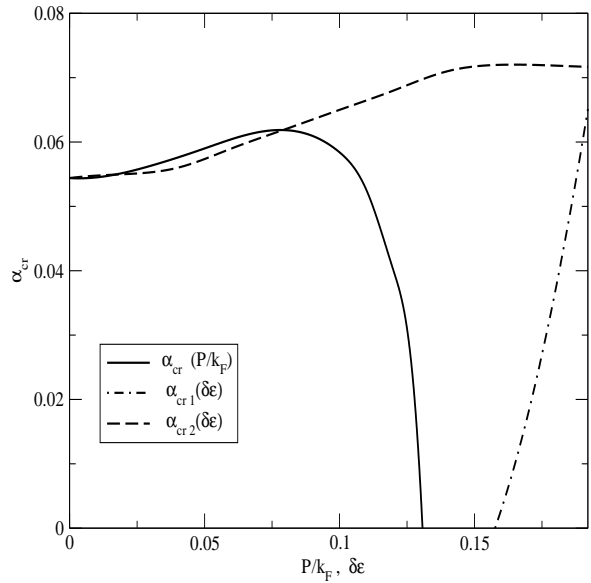}
\vskip 0.5cm
\caption{
  The dependence of the critical asymmetry $\alpha_c$ of the transition from 
  the superfluid to the normal state on the total momentum in the 
  LOFF phase and the deformation parameter in the DFS phase. The 
  model parameters are as in Fig.~1. 
  }\label{MSfig:fig3}
\end{figure}

To elucidate the dominance of the phases with broken-space symmetries
over the asymmetric BCS state consider the modifications in the 
single-particle spectra implied by these phases.
In the asymmetric BCS state, the antisymmetric 
part of the quasiparticle spectrum (\ref{SPECTRUM}), $E_A$, acts in 
the gap equation (\ref{GAP2}) to reduce the phase-space coherence between 
the quasiparticles that pair (when $E_A = 0$ the BCS limit is recovered 
with equal occupations for both particles and perfectly matching 
Fermi surfaces). This blocking effect is responsible for the reduction 
of the gap with increasing asymmetry and its disappearance above 
$\alpha \simeq 0.055$.

When the pairs move with a finite total momentum 
or the Fermi surfaces are deformed, the antisymmetric part of the spectrum,
$E_A$, is modulated with the cosine of the polar angle 
$x$ (in the frame where the $z$ axis is along the symmetry-breaking axis). 
In the case of the plane-wave LOFF phase $E_A\propto x$ while in the
DFS phase  $E_A\propto x^2$. This variation 
acts to restore the phase-space coherence for some values of $x$ 
at the cost of even lesser (than in the BCS phase)
coherence for the remaining values. The effect can be seen explicitely 
in Fig.~\ref{MSfig:fig2} which compares the quasiparticle spectra 
$\omega_1$ and $\omega_2$ in the DFS phase 
for combinations of $\alpha$ and $\delta\epsilon$ for two orthogonal 
directions with the BCS result for the symmetric case (black). 
Along the symmetry-breaking axis the energy separation 
between the quasiparticle spectra is considerably smaller than in the 
asymmetric BCS state; in the orthogonal direction, the opposite is the 
case. Compared to the asymmetric BCS
state the phase-space overlap between pairs is increased in the 
first case and decreased in the second. 
The net result, displayed  in Fig.~\ref{MSfig:fig3}, 
is the {\em increase} in the value of the critical asymmetry 
$\alpha_c$ at which the superfluidity vanishes. At large asymmetries 
the DFS phase exhibits the reentrance effect:
the pairing exists only for the deformed state between the 
lower ($\alpha_{{\rm cr}1}$) and upper ($\alpha_{{\rm cr}2}$) 
critical deformations; see Fig.~\ref{MSfig:fig3}. This figure
also shows the largest possible asymmetry supported by the LOFF 
for a given total momentum $P$; 
this asymmetry is maximal at $P/k_F\sim 0.075$.

An important feature of the spectrum of the asymmetric BCS 
state ($\alpha\neq 0$, $\delta\epsilon =  0 =P$) 
is its gapless nature~\cite{GL1,GL2,GL3,GL4,GL5}- i.e., the existence of nodes 
for one (or both) branches of the spectra 
(cf. the gapped BCS spectrum also shown in Fig.~\ref{MSfig:fig2}). 
Gapless excitations affect the dynamical properties of the superfluid state
such as the transport and collective modes and lead to a
number of  peculiarities in the thermodynamics of this state. 
This feature clearly remains
intact for the phases with broken-space symmetries. As seen in 
Fig.~\ref{MSfig:fig2} the spectrum of the DFS phase covers 
a range bounded by the curves with $x = 0$ and  $x = 1$ and features
nodes at which the quasiparticles can be excited by an infinitely small 
external perturbation. 
The macrophysical manifestations of the LOFF and DFS phases 
such as the response to density perturbations or electromagnetic
probes and the thermodynamic functions (heat capacity, etc.) 
would differ from the ordinary BCS  phase due to the nodes 
{\em and anisotropy} of their spectrum. Such an anisotropy can 
be used to discriminate phases with broken-space symmetries 
in the time-of-flight experiments (see Sect.~\ref{Sect.:exp}).
Moreover, the phases with broken-space symmetries feature 
a larger number of Goldstone modes than the asymmetric BCS phase 
because of the breaking of additional global space symmetries.

\begin{figure}
\begin{center}
\epsfig{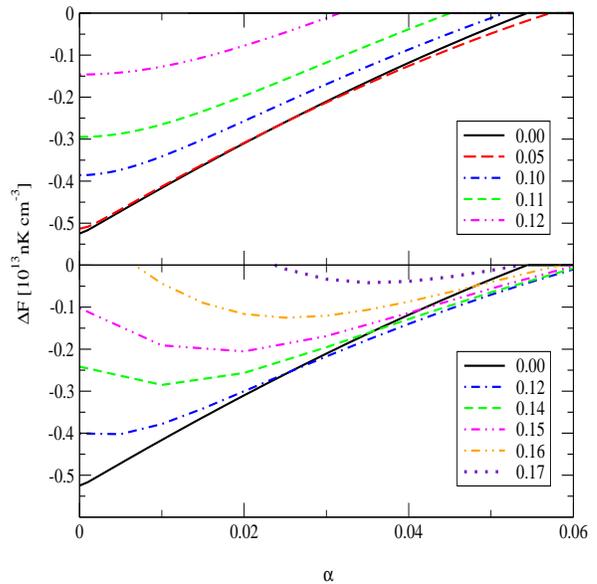}
\end{center}
\caption{(Color online)
The dependence of the free energy of the LOFF (upper panel) 
and the plane-wave DFS phase (lower panel) on the asymmetry 
parameter for several values of the deformation parameter 
$\delta\epsilon$ and the total momentum $P/k_F$  which are indicated 
in the panels. 
}
\label{MSfig:fig4}
\end{figure}
Which phase is the true ground state at a give density asymmetry is 
decided by a comparison of the free energy difference 
$\Delta F = F_S - F_N$  of these phases which are displayed in 
Fig.~\ref{MSfig:fig4}. The LOFF phase is preferred to the normal and 
homogeneous BCS phases in a narrow window of asymmetries,
$0.04 \le \alpha\le 0.057$, and for the total momentum of the pairs,
$P/k_F\sim 0.05$. This is consistent with the results obtained 
in the scheme where the density asymmetry is described in terms 
of the difference in the chemical potentials of the species $\delta\mu$
[the critical value for the BCS phase is $\delta\mu_c^{BCS} = 0.707\Delta(0)$,
while for the LOFF phase  $\delta\mu_c^{LOFF} = 0.755\Delta(0)$, 
where $\Delta(0) \equiv \Delta(\delta\mu = 0)$]
\cite{LARKIN,FULDE}. Note that while 
there is a nontrivial solution to the gap equation for 
 $P/k_F\ge 0.01$, the gain in the pairing energy is 
less than the loss in kinetic energy due to the motion of the condensate 
and the net energy of the LOFF phase is larger than that 
of the asymmetric BCS phase. However, the pairing energy 
of the LOFF phase can be increased by choosing a more complex form 
of the order parameter - e.~g., by keeping a larger number of terms in the
expansion of the order parameter in the Fourier series.

The DFS phase is the ground state of the system 
(i.e., it has a lower free energy than the normal, BCS, and LOFF phases)
in a wider range of asymmetries, $0.03\le \alpha \le 0.06$,  for 
the deformation parameters in the range $0.12\le \delta\epsilon\le 0.16$.
For even larger deformations the gain in the pairing energy does
not compensate the loss in the kinetic energy due to the stretching
of the Fermi surface into ellipsoidal form. Note that the 
free energy is also affected by the reentrance 
effect (i.~e. restoration of pairing correlations as the asymmetry 
is increased). However, while beyond  the reentrance point the 
DFS phase becomes preferable to the unpaired state,
its free energy is still larger than in the homogenous asymmetric 
BCS state. Only at the large asymmetries quoted above does
it become the ground state of the system.

To summarize, the coherence is restored and the strength of
pair correlations is increased in the LOFF phase due to the 
finite momentum of the Cooper pairs. In 
the DFS phase the same is achieve by stretching the spherical Fermi surfaces 
into ellipsoids. The fundamental difference between these
phases is that the translational symmetry remains intact for the 
DFS phase, which breaks only the rotational symmetry, while the LOFF
phase breaks both symmetries. Quantitatively, 
the maximal value of the gap and the absolute value of 
the ground-state free energy is larger in the DFS phase
than in the LOFF phase for asymmetries larger than  $\alpha \simeq 0.04$.
For these asymmetries  both phases are 
favorable over the homogeneous BCS phase. 
However, one should keep in mind that the LOFF phase admits a variety
of lattice forms and the plane-wave structure need not be
the most favored one~\cite{QCD_LOFF}.
\begin{figure}[t]
\epsfig{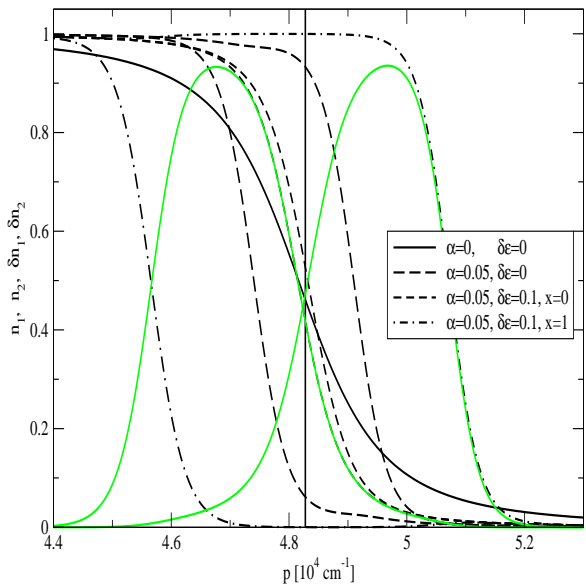}
\caption{
(Color online)
The dependence of the occupation probabilities of two hyperfine 
states on the momentum. The Fermi momentum $k_F = 4.83 \times 10^4$
cm$^{-1}$ is indicated by the vertical line. The labeling of 
the lines is as follows: $\alpha = 0= \delta\epsilon$ (solid line),  
$\alpha = 0.05$ and $\delta\epsilon = 0$ (long-dashed lines),  
$\alpha = 0.05$, $\delta\epsilon = 0.1$, $x = 0$ (short-dashed lines),
and $x = 1$ (dashed-dotted lines).
The bell-shaped light curves show the anisotropy 
-- the difference between the $x=1$ and $x=0$ occupation numbers --
for $\alpha = 0.05$, $\delta\epsilon = 0.1$.
The remaining parameters are as in Fig.~1.
}
\label{MSfig:fig5}
\end{figure}

\section{Detecting the DFS phase in experiments}
\label{Sect.:exp}

The large number of side effects in solid systems (such as the presence
of the ionic lattice or defects) have precluded up to now a clear 
detection of a superfluid phase with broken-space symmetries in these systems.
The situation can be much more favorable in ultracold gases
due to the high control over the experimental conditions.
Recently, Mizushima {\it et al.} proposed a way to detect the
LOFF phase in these systems by imaging the density profiles by 
local spin measurements~\cite{MIZUSHIMA}. 
Below we propose  simple method to detect the DFS phase 
in a Fermi system with rotational symmetry -  e.g., a system which is 
either homogeneous or in a spherical trap. Indeed,
experimental evidence for the phases with broken-space symmetries
can be obtained from studies of their momentum distributions
which, unlike in the homogeneous phase, must be anisotropic in space.
Figure~\ref{MSfig:fig5} shows the occupation numbers in the BCS, 
the asymmetric BCS, and the DFS phases; varying the cosine of the 
polar angle $x$ covers a range of occupation probabilities which includes 
the undeformed asymmetric state. The bell-shaped curves show 
the angular polarization of the occupation numbers in the DFS phase 
defined as $\delta n_{\sigma} = \vert n_{\sigma}(x=1) - n_{\sigma}(x=0)\vert$.
The maximal anisotropy in the occupation 
probabilities of particles along and orthogonal to the symmetry 
breaking axis is about $90\%$. Thus, a direct way
to detect the DFS phase is the measurement of 
the anisotropy in the momentum distribution of the trapped atoms. 
Such a measurement can be realized by the time-of-flight technique
\cite{MBEC1,MBEC2,MBEC3,MBEC4,MBEC5,MBEC6,MBEC7,JILA04}.
This method uses the fact that after releasing the trap,
the atoms fly out freely and an image of their spatial distribution
taken after some time of flight  provides information on 
their momentum distribution when confined inside the trap. 
Assuming that the system was in the deformed superfluid state
one would  detect a mean momentum of
particles of type $1$ (majority) in the direction of symmetry 
breaking by about 90\% larger than that of particles of
type $2$ (minority) in the same direction.
Therefore, the presence of this anisotropy in the 
detected momentum distributions would be evidence for a 
deformed {\em superfluid} state
being the ground-state of the system, as deformation alone
({\em i.~e.} without pairing) would not lower the energy so as to
produce a deformed nonsuperfluid ground-state.
Note that this argument is equally valid for a homogeneous system or for 
an atomic gas in a spherical trap, where no preferred 
direction is introduced by the trapping potential.
For a nonspherical trap, the momentum distributions 
of both species are expected to be deformed {\em in the same way};
therefore, the detection of an anisotropy in momentum distributions
superimposed on the deformation induced by the deformed trap
could be a signature of the DFS phase. Case studies for certain forms 
of deformed trapped (e.g., in the local density approximation) 
would be necessary to quantify the effect of combined deformation due 
to the trap and pairing. 

The direction of spontaneous symmetry-breaking (in $k$ space and,
therefore, also in real space) is chosen by the system randomly
and needs to be located in an experiment to obtain maximum anisotropy.
A clear distinction between the DFS and LOFF phases can be achieved 
in time-of-flight experiments, since the latter predicts 
periodic momentum distributions, independent of the detailed spatial
structure of the phase that nucleates in the ground-state.

\acknowledgments
This work was supported by the 
DGICYT Grant Nos. BFM2002-01868 and HA2001-0023 [Acciones
Integradas] (Spain), the Deutsche  Akademische Austausch Dienst 
and the Sonderforschungsberich 381 of the Deutsche Forschungsgemeinschaft 
(Germany), and the Generalitat de Catalunya (J.M.-P.)


\begin{thebibliography}{0}
\bibitem{COMMENT}{At the relevant temperatures (in the nanokelvin range)
the pairing in $L\neq0$ orbital angular momentum is less favored, 
unless one deals with the pairing among indistinguishable particles 
where the relative $s$-wave pairing is forbidden by the Pauli 
prinicple}.
\end{thebibliography}
\end{document}